\documentclass[aps,titlepage,12pt,superscriptaddress]{revtex4}
\usepackage{epsfig}
\usepackage{times}
\usepackage{color}
\begin{document}

\title{Knowing the past improves cooperation in the future}

\author{Zsuzsa Danku}
\affiliation{Institute of Technical Physics and Materials Science, Centre for Energy Research, Hungarian Academy of Sciences, P.O. Box 49, H-1525 Budapest, Hungary}

\author{Matja{\v z} Perc}
\email{matjaz.perc@uni-mb.si}
\affiliation{Faculty of Natural Sciences and Mathematics, University of Maribor, Koro{\v s}ka cesta 160, SI-2000 Maribor, Slovenia}
\affiliation{School of Electronic and Information Engineering, Beihang University, Beijing 100191, China}
\affiliation{Complexity Science Hub Vienna, Josefst\"{a}dterstra{\ss}e 39, A-1080 Vienna, Austria}

\author{Attila Szolnoki}
\email{szolnoki.attila@energia.mta.hu}
\affiliation{Institute of Technical Physics and Materials Science, Centre for Energy Research, Hungarian Academy of Sciences, P.O. Box 49, H-1525 Budapest, Hungary}

\begin{abstract}\noindent
\\Cooperation is the cornerstone of human evolutionary success. Like no other species, we champion the sacrifice of personal benefits for the common good, and we work together to achieve what we are unable to achieve alone. Knowledge and information from past generations is thereby often instrumental in ensuring we keep cooperating rather than deteriorating to less productive ways of coexistence. Here we present a mathematical model based on evolutionary game theory that shows how using the past as the benchmark for evolutionary success, rather than just current performance, significantly improves cooperation in the future. Interestingly, the details of just how the past is taken into account play only second-order importance, whether it be a weighted average of past payoffs or just a single payoff value from the past. Cooperation is promoted because information from the past disables fast invasions of defectors, thus enhancing the long-term benefits of cooperative behavior.
\end{abstract}

\maketitle

\noindent Our exceptional other-regarding abilities originate from our struggles to withstand the evolutionary pressure during the nascent period of the genus \textit{Homo}. Through alloparental care and provisioning for the children of others, cooperation enabled us to rear offspring that survived \cite{hrdy_11}. Cooperation has also enhanced in-group solidarity, which ultimately helped us to mitigate between-group conflicts in our earliest societies \cite{bowles_11}. It is fascinating to grasp how our humble beginnings, where cooperation was primarily a necessity for us to survive, led to the flourishing human societies that dominate the planet today. We are indeed super-cooperators \cite{nowak_11}, enjoying the harvest of collective efforts on an unprecedented scale. Undoubtedly, cooperation is at the heart of the main evolutionary transitions that led from single-cell organisms to complex animal and human societies \cite{maynard_95}.

But to cooperate is costly, and the act benefits others. As such, cooperation ought to be unsustainable according to Darwin's \textit{The Origin of Species}. If only the fittest survive, why should we care for and contribute to the public good if freeriders can enjoy the same benefits for free? Research based on evolutionary game theory \cite{hofbauer_98, nowak_06,javarone_18,tanimoto_15} has revealed key mechanisms that explain the evolution of cooperation, including kin selection, direct reciprocity, indirect reciprocity, network reciprocity and group selection \cite{nowak_s06}. Cooperation can also be promoted with positive and negative incentives \cite{andreoni_aer03, rand_tcs13, yamagishi_pnas12, weber_nc18, perc_pr17, hilbe_nhb18, tanimoto_pre14}, including rewards for behaving prosocially \cite{dreber_n08, wu_y_srep17, hilbe_prsb10, szolnoki_njp12, szolnoki_prsb15} and punishment for freeriding \cite{fehr_aer00, boyd_pnas03, henrich_s06b, hauser_jtb14, szolnoki_prx13, gao_l_srep15, cong_r_srep17, chen_pcbi18, takesue_epl18, liu_srep17, liu_jz_csf18, szolnoki_prx17}.

However, just like cooperation incurs a cost for the benefit of others, so does the provisioning of rewards and sanctions. Individuals that abstain from dispensing such incentives therefore become second-order freeriders \cite{fehr_n04}, and the puzzle of cooperation is frequently not solved but just diverted to another level. Strategy-neutral mechanisms that promote cooperation do not have this drawback, although they have the downside that it is often unforeseeable how the competing strategies will be affected. An important example is the positive effect of population heterogeneity on the evolution of cooperation. The heterogeneity may manifest in the diverse number of neighbors an individual has in a social network \cite{santos_prl05}, in differences of the microscopic dynamics that governs strategy changes \cite{szolnoki_epl07}, it can be cast as social diversity \cite{perc_pre08, santos_n08}, or it can manifest as a public resource that changes over time and depends on the strategic choices of individuals \cite{hilbe_n18}. Coevolutionary strategy-neutral rules have also been presented that can enhance cooperation \cite{perc_bs10}. For example, when aging adversely affects reproduction this has a highly selective impact on the propagation of cooperators and defector in favor of the former \cite{szolnoki_pre09}.

Here we significantly expand the scope of strategy-neutral rules that promote cooperation, in particular by taking into account the role of time and information from the past to inform actions in the future. This important aspect of evolutionary dynamics has recently been studied by Hauser et al. \cite{hauser_n14}, who noted that the overexploitation of renewable resources today has a high cost on the welfare of future generations, and moreover, that future generations cannot reciprocate actions made today. A new experimental paradigm was proposed, where a line-up of successive generations can each either extract a resource to exhaustion or leave something for the next group. Research revealed that exhausting the resource maximizes the payoff for the present generation, but leaves all future generations empty-handed. However, the tragedy of the commons could be averted if the exploitation is decided democratically by a vote.

In what follows, we propose a simple mathematical model that builds on the formalism of evolutionary social dilemmas, where past payoffs are taken into account to inform strategies in the future. We consider a weighted moving average over a period of past payoffs, as well as individual chosen payoff from the past, as determinants of the current fitness of a player that decides its future strategy. We find that such minimal interventions suffice to significantly change the course of evolution in favor of cooperation in social dilemmas. Indeed, simply knowing the past and taking it into account improves cooperation in the future. Thereby no assumptions need to be made as to who or which strategy has this information, and the implementation of how the past is taken into account also plays only second-order importance. As we will show, cooperation is promoted because the strategy-neutral rule has a highly asymmetric effect on the evolution of the two competing strategies. While the invasion of defectors into cooperative clusters is strongly decelerated, cooperative domains continue to grow, albeit slowly, which ultimately reveals and amplifies the long-term benefits of cooperation.

\section*{Results}

\subsection*{Mathematical model}

\noindent We build on the traditional social dilemma model, where players can choose either to cooperate or defect. Mutual cooperation yields the reward $R$, mutual defection leads to punishment $P$, and the mixed choice gives the cooperator the sucker's payoff $S$ and the defector the temptation $T$. By setting $R = 1$ and $P=0$ as fixed, the remaining two payoffs occupy $-1 \leq S \leq 1$ and $0 \leq T \leq 2$, where if $T>R>P>S$ we have the prisoner's dilemma game, $T>R>S>P$ yields the snowdrift game, and $R>T>P>S$ the stag-hunt game. Players play the social dilemma in a pairwise manner (see Methods for details), whereby player $i$ at instance $n$ of the game obtains the payoff $P_{n,i}$.

In the first place, we consider a weighted moving average over past payoffs, such that the further back in time, the lesser the weight given to a particular payoff. Formally, the final payoff of player $i$ is then
\begin{equation}
P_i = \frac{P_{0,i}+\sum_{m=1}^{M} \alpha^m P_{m,i}}{1+\sum_{m=1}^{M} \alpha^m} \, ,
\end{equation}
where $P_{m,i}$ is the payoff of player $i$ that was collected $m$ rounds back in time from the present. Moreover, $\alpha$ is a free parameter that determines how fast the weight factor decays for increasing values of $m$, and it also determines, albeit indirectly, the memory length $M$. In particular, the memory cutoff occurs when the weight factor goes below the $0.01$ threshold. It is worth pointing out that for $\alpha=0$ this model reverts back to the traditional social dilemma where the past is not considered. On the other hand, in the $\alpha \to 1$ limit the memory window is extended to all previous payoffs, but the normalization still ensures finite payoff values, and thus an ongoing evolutionary dynamics without locally frozen states.

Secondly, as an alternative to the above-described mathematical model, we also consider the variant where a single payoff from the past is used instead of the current payoff for determining the strategy change probability (see Methods for the later). However, we retain the argument that the further the payoff back in time, the lesser its impact ought to be, and thus the lower the probability that it will be chosen instead of the current payoff. Formally, at instance $n$ of the game, instead of $P_{n,i}$ we thus consider one chosen earlier payoff value $P_\tau$ of the same player $i$ with the probability $\nu=\exp(-\tau/s)$, where $s=-100/\ln(0.01)$ simply determines a natural time decay such that the chance to use $P_{100}$ ($\tau=100$ steps in the past) is only 1 \%. Alternatively, the present payoff value $P_{n,i}$ is used with probability $1-\nu$. Similarly to the previously introduced model, in this case at $\tau=0$ and in the $\tau \to \infty$ limit the model reverts back to the traditional social dilemma where the past is not considered.

As we will show next, both variants of the mathematical model, although significantly different per definition, yield very similar evolutionary outcomes. Cooperation is strongly promoted in both cases, and this is due to the same microscopic mechanism. We will show that the invasion of defectors into cooperative clusters is strongly decelerated, whilst cooperative domains continue to grow slowly but steadily. Ultimately, this biased effect of a strategy-neutral intervention in the form of taking into account past payoffs reveals and amplifies the long-term benefits of cooperation. Interestingly, it matters not how the past is taken into account, whether by means of a weighted average of past payoffs or just a single payoff value from the past, thus revealing a universally valid mechanism for cooperation in social dilemmas.

\subsection*{Evolution of cooperation}

\begin{figure}
\centerline{\epsfig{file=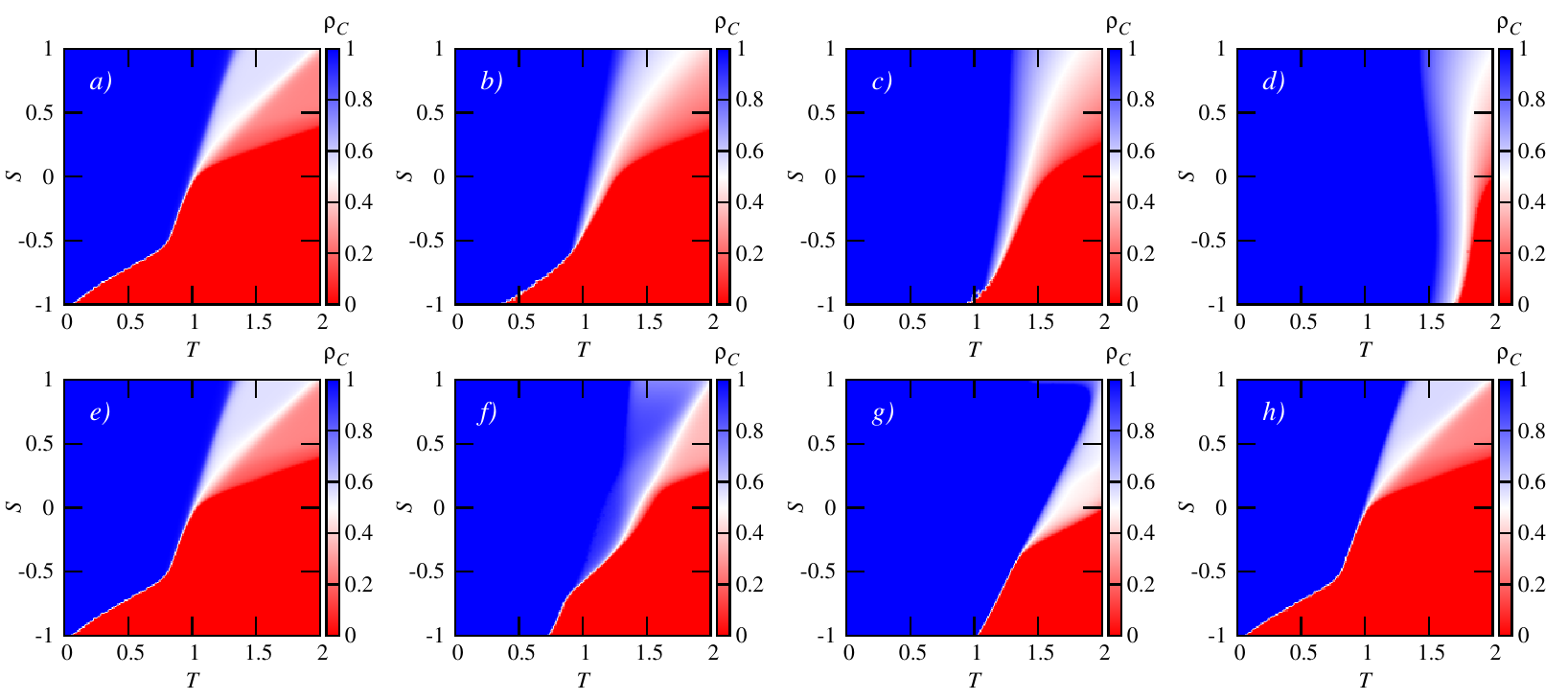,width=16.0cm}}
\caption{Heat maps of cooperation reveal that taking into account the past improves cooperation in the future. The color encodes the stationary fraction of cooperators $\rho_C$, as indicated by the color bars. Upper row shows results obtained with the mathematical model with a weighted moving average over past payoffs. From (a) to (d) the values of the decay parameter $\alpha$ are $0$, $0.5$, $0.8$ and $0.95$, respectively. Lower row shows results obtained with the mathematical model where a single past payoff value is considered with a weighted probability. From (e) to (h) the values of the time delay $\tau$ are $0$, $1$, $3$ and $50$, respectively.}
\label{heatmaps}
\end{figure}

To highlight the conceptual similarities between the two seemingly very different variants of the mathematical model, we present the obtained results in parallel. The upper row of Fig.~\ref{heatmaps} shows how the stationary fraction of cooperators varies in the $T-S$ plane for the model with a weighted moving average over past payoffs. Results for four representative values of the decay parameter $\alpha$ are presented, whereby it can be observed that the longer the memory window into the past, the more the cooperators dominate even in the most challenging prisoner's dilemma quadrant ($T>1$ and $S<0$). This observation is also in agreement with the evolutionary outcome of conceptually similar models that have been studied in the past \cite{liu_yk_pa10, wang_xw_pla16,javarone_epjb16}.

\begin{figure}
\centerline{\epsfig{file=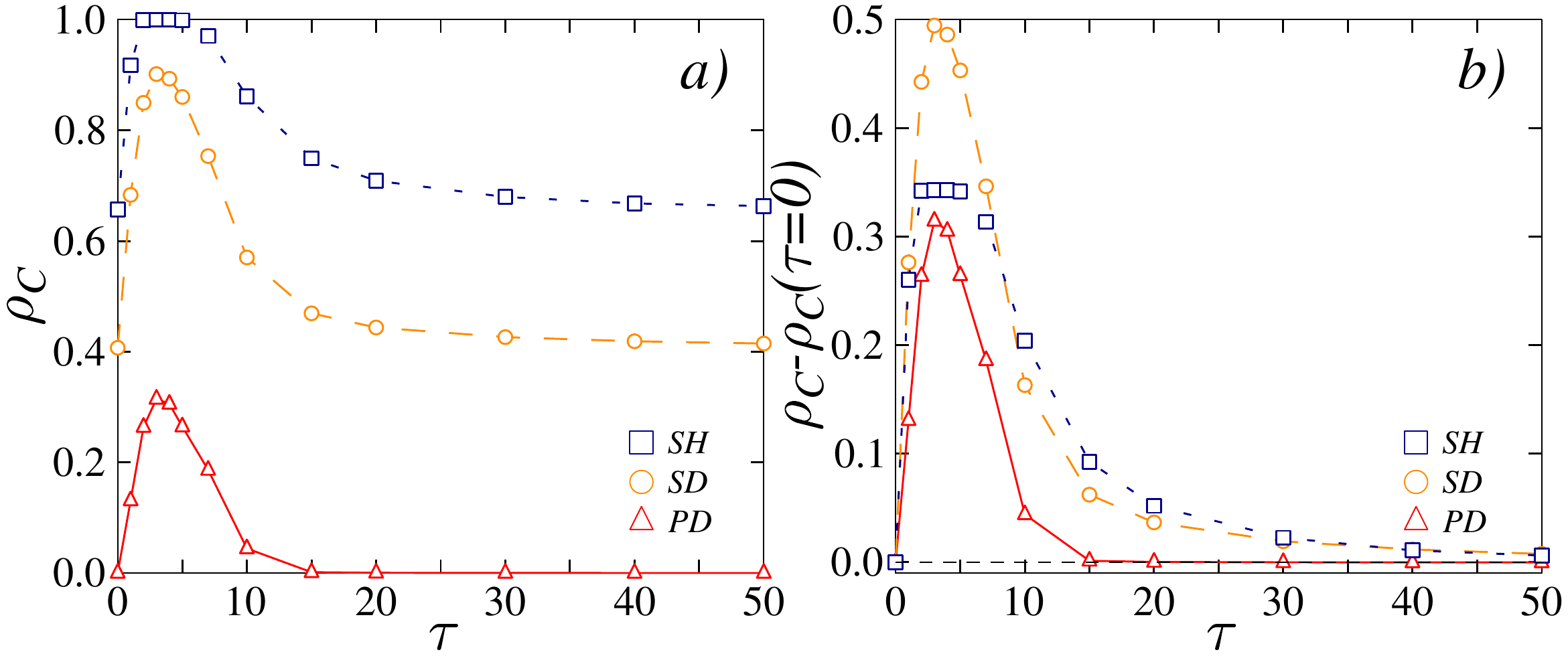,width=12cm}}
\caption{Average cooperation levels for different social dilemma games reveal an optimal value of the time delay $\tau$ at which cooperation thrives best. The legend in both panels indicates different social dilemma types (SH $=$ stag-hunt, SD $=$ snowdrift, PD $=$ prisoner's dilemma). (a) The average cooperation level, obtained by averaging over all $(T,S)$ pairs that correspond to a particular social dilemma, in dependence on the time delay $\tau$. (b) The difference between the average cooperation level and the average cooperation level obtained at $\tau=0$ in dependence on the time delay $\tau$. It can be observed that, relatively, the snowdrift $(T,S)$ quadrant benefits the most in terms of cooperation promotion.}
\label{taudep}
\end{figure}

In comparison, the lower row of Fig.~\ref{heatmaps} shows the same results, but for the model where a single past payoff value is considered with a weighted probability. Here the improvement towards more cooperation is also visible as the time delay $\tau$ increases from left to right, although it is not monotonous as in the upper row. In particular, since large values of $\tau$ make it increasingly unlikely that a past payoff will be considered instead of the present one (see model definition), an intermediate value of $\tau$ is in fact optimal for the evolution of cooperation (panel g in the lower row of Fig.~\ref{heatmaps}). To further clarify the dependence of cooperation on $\tau$, we show in Fig.~\ref{taudep}(a) the average level of cooperation over all $(T,S)$ pairs that determine the same social dilemma type, as well as the difference with the $\tau=0$ case in panel (b). The optimal intermediate value of $\tau$ is clearly inferable for all three different social dilemma types (see figure legend). Moreover, it can be observed that the optimal value of $\tau$ is the same for all social dilemma types, and that relatively to the $\tau=0$ baseline case the snowdrift quadrant benefits the most (on average).

\begin{figure}
\centerline{\epsfig{file=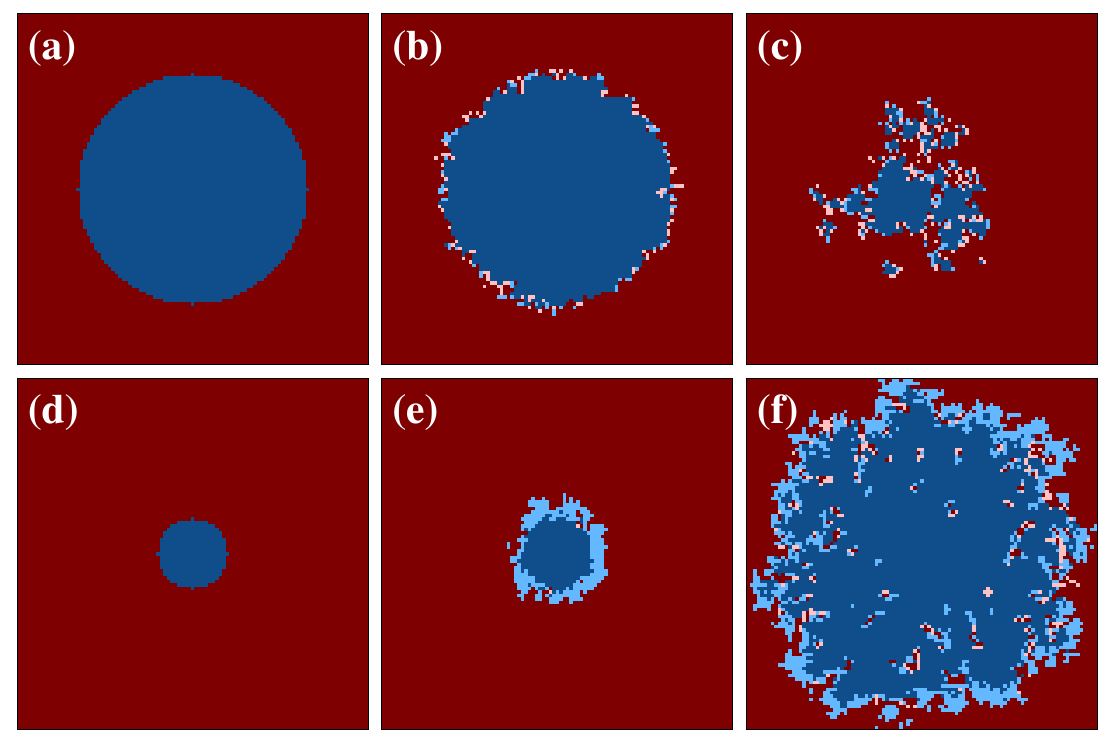,width=15cm}}
\caption{Representative spatial evolutions of the two competing strategies, as obtained with the mathematical model with a weighted moving average over past payoffs. The time increases from left (initial state) to right. Weak (strong) cooperators are depicted light (dark) blue, while weak (strong) defectors are depicted light (dark) red. Strategies are considered strong (weak) if the current strategy of a players is the same (different) as its strategy in the middle of the time window used for the moving average. Upper row depicts snapshots of the square lattice for $T=1.3$, $S=-0.1$ and $\alpha=0.5$. We have used a prepared initial state (a sizable round cooperative domain surrounded by defectors) and a small $101 \times 101$ lattice size for clarity. In this case the population ultimately evolves towards a full defector state (not shown). Lower row depicts snapshots of the square lattice for $T=1.3$, $S=-0.1$ and $\alpha=0.9$. The coloring and other details are the same as in the upper row. Here a smaller round cooperative domain evolves towards a fully cooperative final state (not show). Thus, just an increase in the value of $\alpha$ from $0.5$ to $0.9$ completely changes the evolutionary outcome in this case.}
\label{snapalpha}
\end{figure}

Next, we present representative spatial evolutions of the two competing strategies, first for the mathematical model with a weighted moving average over past payoffs in Fig.~\ref{snapalpha}. The goal is to understand the microscopic mechanism that is responsible for the above summarized cooperator-supporting effects. To that effect, we use a special coloring technique where we distinguish weak and strong cooperators as well as weak and strong defectors. The distinction between weak and strong is based on whether the current strategy of a player agrees with its strategy in the past (we use the middle of the time window for the weighted moving average, or simply the strategy at current time minus $\tau$). Accordingly, strong (weak) cooperators are denoted by dark (light) blue, while strong (weak) defectors are denoted by dark (light) red.

We compare outcome obtained for $T=1.3$, $S=-0.1$ and $\alpha=0.5$ in the upper row and $\alpha=0.9$ in the lower row. As the upper row of Fig.~\ref{snapalpha} illustrates, even if we use special initial conditions in the form of a sizable cooperative domain, if the time window for the moving average is too short cooperators can not escape extinction (final state is not show, but it can be observed in the animation \cite{1_a05_movie}). Although it seems that strong cooperators (dark blue) can initially beat strong and weak defectors (dark and light red), the reality soon transpires, and it is due to the fact that weak defectors are not really weak. Since the time window in the past is short, current payoffs have significant weight, and hence the population behaves as if the past is basically not taken into account. Consequently, the high $T$ and low $S$ value provide too large of an advantage for defectors, who are therefore ultimately wipe out all cooperators.

However, if we prolong the width of the time window into the past by using $\alpha=0.9$, the lower row of Fig.~\ref{snapalpha} shows that in this case the impending full defection state can be reverted into a full cooperation state. We also provide an animation corresponding to these snapshots in \cite{1_a09_movie}. In this case the time-averaged payoffs provide an efficient support for cooperators, such that defectors, who can only enjoy a temporarily high payoff, experience a strongly decelerated invasion. Moreover, a patch composed solely of defectors becomes especially sensitive because strong cooperators (dark blue) can invade them successfully, thus leaving weak cooperators (light blue) in their wake. But weak cooperators are not really weak because they can still enjoy the long-term benefits of being surrounded by other cooperators. Defectors surrounded by other defectors enjoy no such benefits. Therefore weak cooperators can easily invade strong defectors (dark red). Furthermore, weak cooperators gradually become strong cooperators over time, and in this way it can be observed that dark blue eventually invade dark red domains by using the light blue players as a shield in front of them. It is worth pointing out that such protective layers can emerge in rather different systems \cite{szolnoki_pre16b, szolnoki_prx17}, which thus underlines that the observed pattern formation is universally applicable under appropriate conditions.

\begin{figure}
\centerline{\epsfig{file=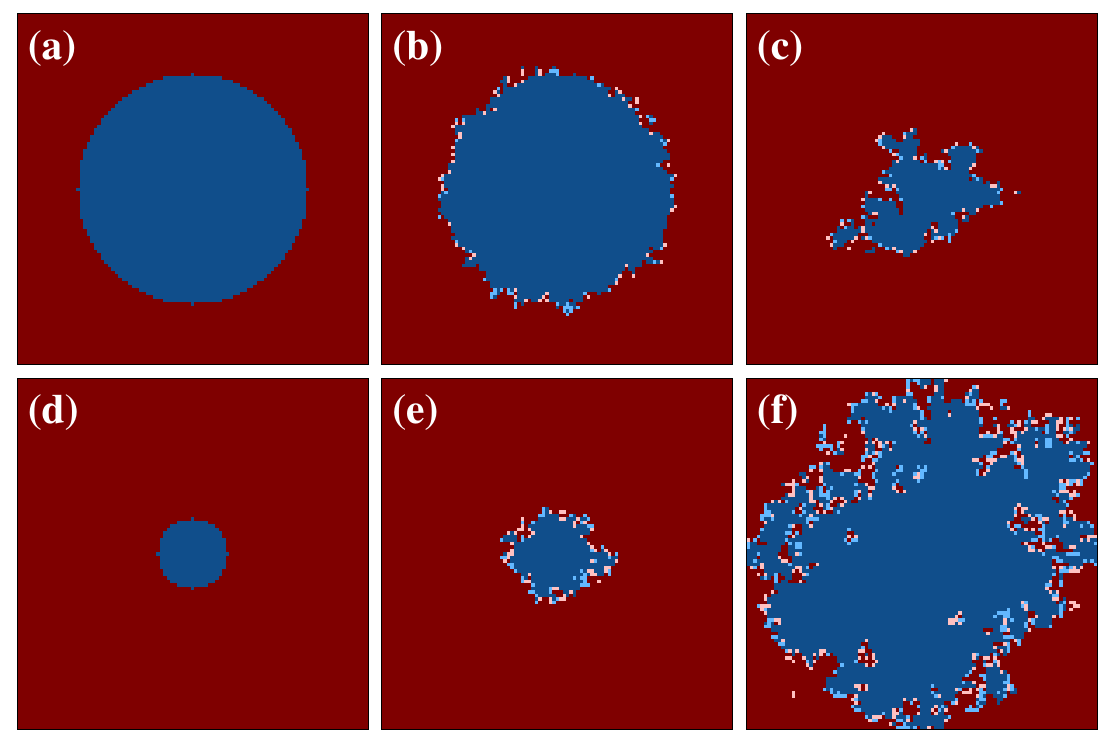,width=15cm}}
\caption{Representative spatial evolutions of the two competing strategies, as obtained with the mathematical model where a single past payoff value is considered with a weighted probability. The coloring and other details are the same as in Fig.~\ref{snapalpha}. Here strategies are considered strong (weak) if the current strategy of a players is the same (different) as its strategy at current time minus $\tau$. Upper row depicts snapshots of the square lattice for $T=1.3$, $S=-0.4$ and $\tau=1$. In this case the population ultimately evolves towards a full defector state (not shown). Lower row depicts snapshots of the square lattice for $T=1.3$, $S=-0.4$ and $\tau=3$. Here a smaller round cooperative domain evolves towards a fully cooperative final state (not show). Thus, just an increase in the value of $\tau$ from $1$ to $3$ completely changes the evolutionary outcome in this case. We emphasize that, although the two considered mathematical models are significantly different per definition, the spatiotemporal evolutionary dynamics is strikingly similar (compare with Fig.~\ref{snapalpha}).}
\label{snaptau}
\end{figure}

Indeed, by looking at representative spatial evolutions of the two competing strategies as obtained for the mathematical model where a single past payoff value is considered with a weighted probability in Fig.~\ref{snaptau}, the similarity with the snapshots shown in Fig.~\ref{heatmaps} is quite striking. In the upper row, we use $T=1.3$, $S=-0.4$ and $\tau=1$, in which case we observe that such a short time delay does not really help cooperators. Defectors soon rise to complete dominance due to the high $T$ and low $S$ value, as can be observed also in the corresponding animation that we provide in \cite{2_t1_movie}. But by using $\tau=3$ instead of $\tau=1$, the unfortunate evolutionary outcome is completely reversed. As shown in the lower row of Fig.~\ref{snaptau} (see also the animation \cite{2_t3_movie}), practically the same spatiotemporal dynamics is in place as described above for the lower row of Fig.~\ref{snapalpha}. Naturally, the protective light blue belt that is made up of weak cooperators is not as thick because picking up a single payoff value from the past cannot provide quite as firm support as a weighted moving average over many past payoffs. Nevertheless, we witness basically the same mechanism. Defectors cannot utilize their actual advantage stemming from the high $T$ and low $S$ value near cooperators, which effectively prohibits them to invade cooperative domains, whilst the latter grow slowly but steadily until defectors die out and the long-term benefits of cooperation are fully revealed.

Finally, to provide quantitative support for the above-outlined microscopic mechanism and for its similarity in both variants of the considered mathematical model, we measure the elementary steps between different subgroups of strategies when the evolution is launched from a random initial state. For easier reference we denote strong cooperators by $C_C$, weak cooperators by $C_D$, strong defectors by $D_D$, and weak defectors by $D_C$. We monitor how these four groups interact with each other while the system evolves towards the stationary state.

\begin{figure}
\centerline{\epsfig{file=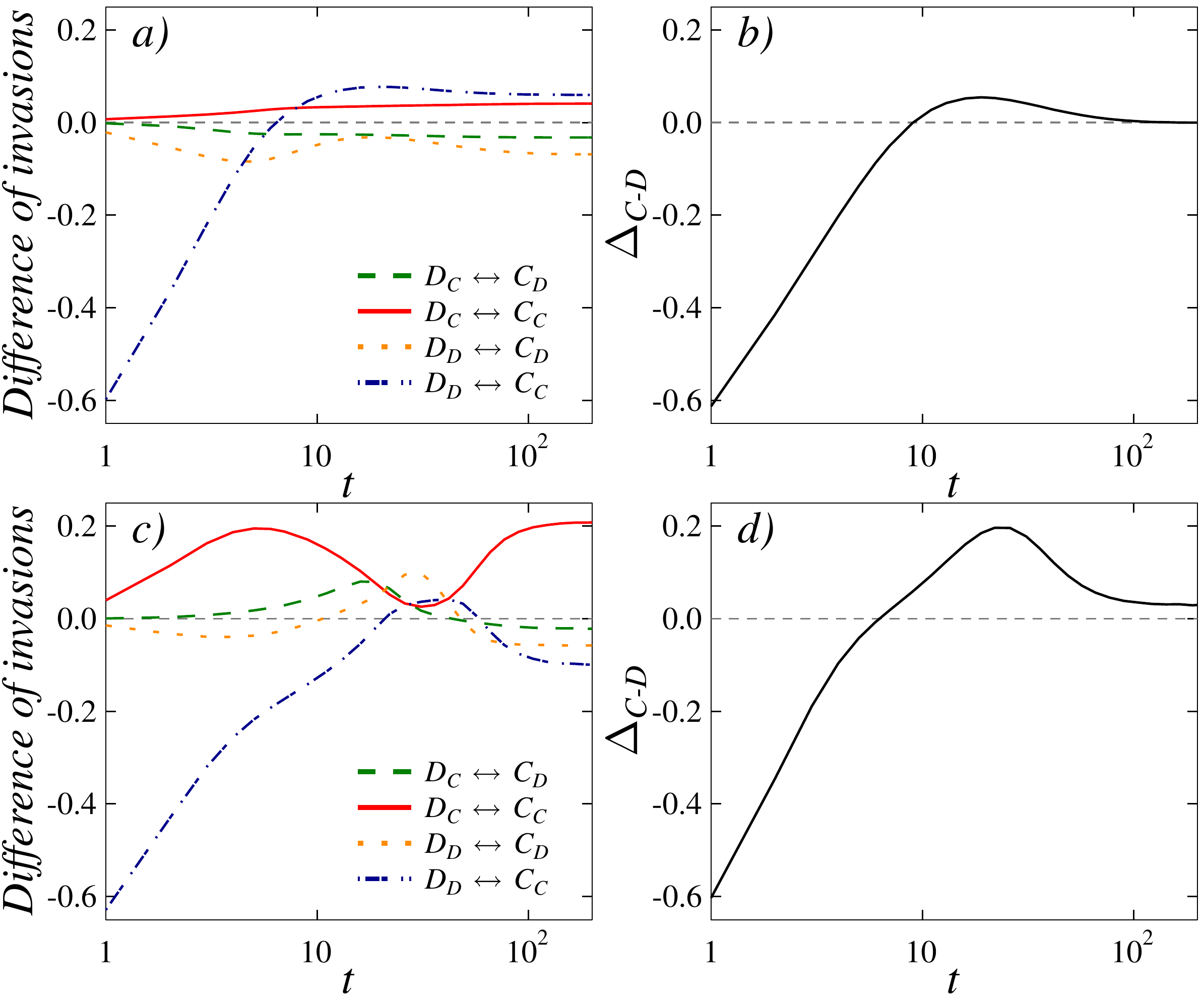,width=9cm}}
\caption{An analysis of invasion rates over time between different subgroups in the mathematical model with a weighted moving average over past payoffs reveals that cooperators benefit on the expense of defectors for sufficiently large values of the decay parameter $\alpha$. The considered subgroups are strong cooperators $C_C$, weak cooperators $C_D$, strong defectors $D_D$, and weak defectors $D_C$ (see legend for monitored invasion rates). The two right panels show accumulated differences in the invasion rates between cooperators and defectors (both strong and weak). Upper row shows results obtained for $T=1.2$, $S=0$ and $\alpha=0.5$ (final state full defection), while the lower row shows results obtained for $T=1.2$, $S=0$ and $\alpha=0.9$ (final state full cooperation). The key difference between $\alpha=0.5$ and $\alpha=0.9$ is the invasion rate difference between weak defectors and strong cooperators ($D_C \leftrightarrow C_C$, denoted by solid red line). For $\alpha=0.9$ this curve is strongly positive during considerably long time spans, and it is this perturbation of the elementary $D_D \leftrightarrow C_C$ process that ultimately tips the balance in favor of cooperators.}
\label{1_inv}
\end{figure}

In Fig.~\ref{1_inv} we show results that correspond to the values of the decay parameter $\alpha$ used in Fig.~\ref{snapalpha}. Notably, there $\alpha=0.5$ (upper row) resulted in full defection, while $\alpha=0.9$ (lower row) resulted in full cooperation. The two panels on the left show the interactions between different strategy subgroups (see legend), and the two panels on the right show the accumulated values. The latter inform us how the fractions of the two competing strategies change over time, whereby a positive value of the difference means that the fraction of cooperators grows on the expense of defectors. By comparing left and right panels, we find that the decisive microscopic process that tips the scale in favor of cooperators in the $\alpha=0.9$ case is the invasion between $D_D$ and $C_C$ groups (note that the dashed-dotted blue curve in the left panel changes simultaneously with the black curve in the right panel). An important difference between $\alpha=0.5$ and $\alpha=0.9$ is that the invasion between weak defectors and strong cooperators ($D_C \leftrightarrow C_C$, denoted by solid red line) in the later case retains a significantly high positive value over long periods of time. And it is the resulting small perturbation of the basic $D_D \leftrightarrow C_C$ process (dash-dotted blue line) that finally paves the way towards cooperator dominance.

\begin{figure}
\centerline{\epsfig{file=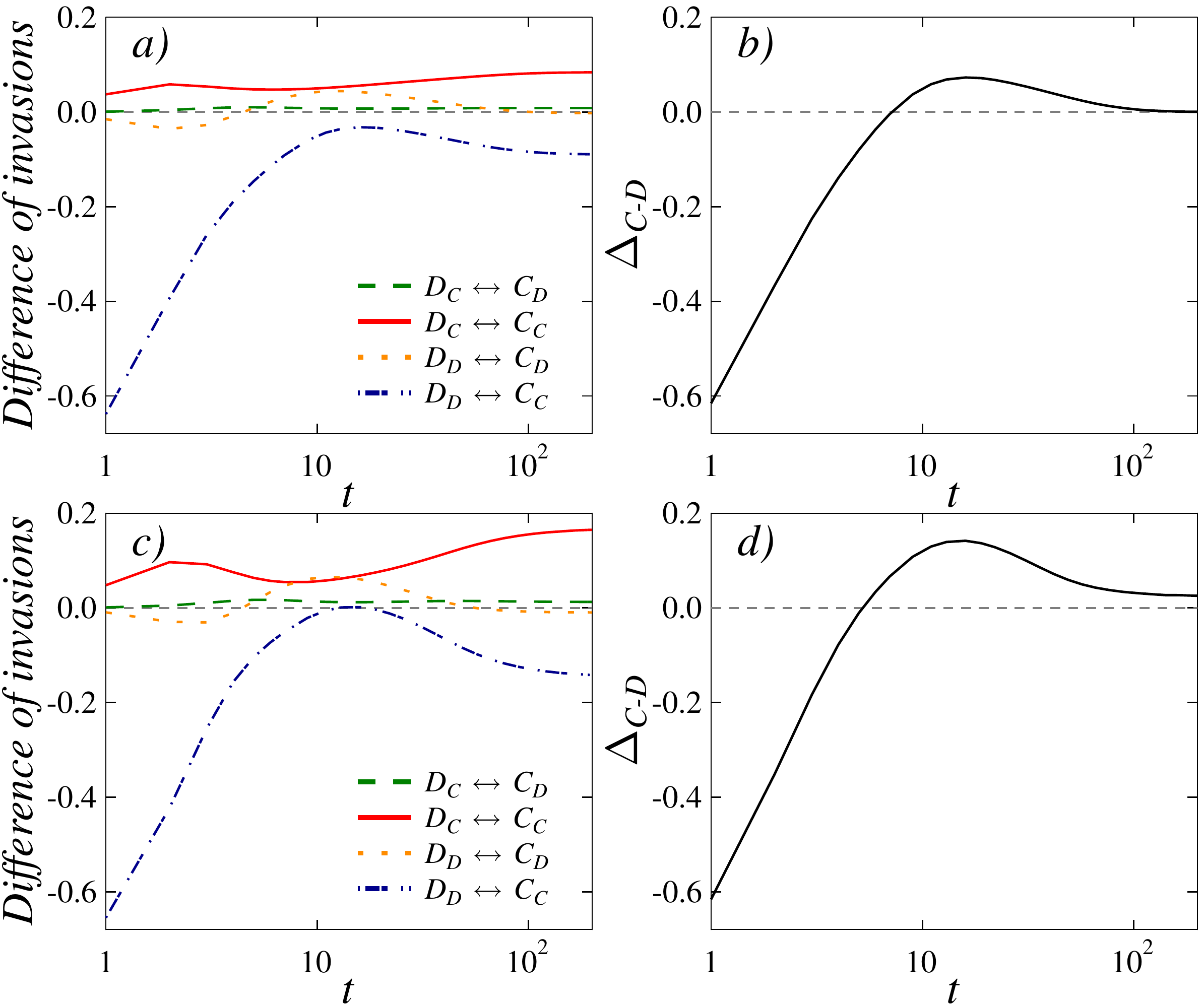,width=9cm}}
\caption{An analysis of invasion rates over time between different subgroups in the mathematical model where a single past payoff value is considered with a weighted probability reveals that cooperators benefit on the expense of defectors for intermediate values of the time delay $\tau$. The considered subgroups are the same as in Fig.~\ref{1_inv} (see legend for monitored invasion rates). Upper row shows results obtained for $T=1.4$, $S=0$ and $\tau=1$ (final state full defection), while the lower row shows results obtained for $T=1.4$, $S=0$ and $\tau=3$ (final state full cooperation). The key difference between $\tau=1$ and $\tau=3$ is, exactly the same as in Fig.~\ref{1_inv}, the invasion rate difference between weak defectors and strong cooperators ($D_C \leftrightarrow C_C$, denoted by solid red line). For $\tau=3$ this curve is strongly positive during considerably long time spans, and it is this perturbation of the elementary $D_D \leftrightarrow C_C$ process that ultimately tips the balance in favor of cooperators. We again highlight that, despite their differences, the two considered mathematical models owe the promotion of cooperation to precisely the same microscopic process, thus revealing in the information from the past a universally valid mechanism for more cooperation in the future.}
\label{2_inv}
\end{figure}

Results presented in Fig.~\ref{2_inv} for the model where a single past payoff value is considered with a weighted probability can be understood along the same lines. Here $\tau=1$ (upper row) results in full defection, while $\tau=3$ (lower row) results in full cooperation (see also Fig.~\ref{snaptau} for the corresponding snapshots). In this case too the invasions between strong defectors and strong cooperators are relevantly affected by the invasions between weak defectors and strong cooperators (dash-dotted blue line and solid red line, respectively), such that for $\tau=3$ cooperators turn out the winners. This quantitative comparison further corroborates the fact that in both considered variants of the studied mathematical model practically the same microscopic mechanism plays the key role in ensuring more favorable evolutionary outcomes. Thus, despite differences in the integration of past payoffs into the model, our research reveals that these details play only second-order importance in ensuring that information from the past is utilized to improve cooperation in the future.

\section*{Discussion}

\noindent We have proposed and studied a mathematical model based on evolutionary game theory that shows how using the past performance as the benchmark for success can significantly improve the evolution of cooperation in social dilemmas. We have considered two variants of the mathematical model, namely one with a weighted moving average over past payoffs, and one where a single past payoff value is considered with a weighted probability. We have shown that, irrespective of the differences in how the past is taken into account, knowing and incorporating it into the evolutionary process can fundamentally change the evolutionary outcome in favor of cooperation. In particular, if the window into the past that is used for the weighted moving average is sufficiently long, or if the single past payoff value is not too old, a full defection state can be reverted into a full cooperation state.

Our research has revealed further that the mechanism that is responsible for the promotion of cooperation does not depend on the details of the model implementation. On the contrary, the observed evolutionary dynamics is universally valid in that it relies on a strongly asymmetric effect the strategy-neutral `taking into account the past' rules (both variants) have on the evolution of the two competing strategies. More precisely, while defectors are adversely affected by a strong deceleration of their ability to invade cooperative clusters, cooperators experience only a mild slowdown in the build up of their domains. The net effect of this asymmetry is that defectors perish while cooperators thrive, even under adverse conditions where normally defectors would long dominate completely.

From the microscopic point of view, the dramatic shift in the evolutionary dynamics is due to the spontaneous formation of a protection shield that is formed by the so-called weak cooperators -- these are cooperators whose current strategy is different from the one in the past. In other words, weak cooperators have managed to arrive to their current strategy even though they were defectors in the considered past, and they pave the way for strong cooperators -- these are cooperators that were not defectors in the considered past -- to successfully invade defectors. We have quantified the emergence of the protection shield by monitoring invasion rates over time between different strategy subgroups, and we have shown that it is the change in the `strong cooperators $\to$ strong defectors' elementary process that ultimately tips the balance in favor of the overall cooperator dominance. As already emphasized, although the two variants of the mathematical model are significantly different per definition, the promotion of cooperation is in both cases due to precisely the same microscopic process, thus revealing in the information from the past a universally valid mechanism for more cooperation in the future.

It is fascinating to learn how minimal and strategy-neutral interventions into established mathematical models of cooperation suffice to revert defection into cooperation. And while it is clear that information from the past can relevantly inform actions in the future -- to repeat the quote by Edmund Burke ``Those who don't know history are doomed to repeat it'' -- experimental and theoretical research on cooperation is only starting to come to grips with all the implications of this fact. A beautiful example of research along these lines was the 2014 paper ``Cooperating with the future'' by Hauser {\it et al.}, where a new experimental paradigm was proposed to preserve resources for future generations based on voting towards a more responsible and moderate extraction in the present time. We hope that our theoretical model will help stem the tide further towards a deeper appreciation of the fact that our actions today and in the past may have far reaching consequence in the future, and that thus theory and experiments critically probing human cooperation should urgently take this into account.

\section*{Methods}

\noindent We have studied evolutionary outcomes of the proposed mathematical model on a square lattice of size $L^2$ with the von Neumann neighborhood and periodic boundary conditions. The square lattice is the simplest of networks that properly describes the fact that the interactions among us are inherently structured rather than random. By using the square lattice, we continue a long-standing tradition that begun with the work of Nowak and May \cite{nowak_n92b}, and which has since emerged as a default setup to reveal all evolutionary outcomes that are feasible in structured populations \cite{perc_pr17}.

Initially, each player $i$ was designated either as a cooperator $(s_i=C)$ or defector $(s_i=D)$ with equal probability. Subsequently, we have applied the Monte Carlo simulation method with the following three elementary steps at each particular time $n$. Firstly, a randomly selected player $i$ acquires its payoff $P_{n,i}$ by playing the game with all its four nearest neighbors. Secondly, one randomly chosen neighbor of player $i$, denoted by $j$, also acquires its payoff $P_{n,j}$ by playing the game with all its four neighbors. Finally, taking the past payoffs into account as described in the Results section, player $i$ with its final payoff $P_i$ adopts the strategy $s_j$ from player $j$ with the probability
\begin{equation}
W = \frac{1}{1+\exp[(P_i-P_j)/K]} \, ,
\end{equation}
where $K$ quantifies the uncertainty by strategy adoptions \cite{szabo_pre98}. In the $K \to 0$ limit, player $i$ copies the strategy of player $j$ if and only if $P_j > P_i$. Conversely, in the $K \to \infty$ limit, payoffs seize to matter and strategies change as per flip of a coin. Between these two extremes players with a higher payoff will be readily imitated, although under-performing strategies may also be adopted, for example due to errors in the decision making or imperfect information. Without loss of generality we have here used $K=0.1$. Repeating the above three elementary steps $L^2$ times constitutes one full Monte Carlo step, which thus gives a chance to every player to change its strategy once on average.

Presented results were obtained on a square lattice of linear size from $L=100$ to $L=800$ to avoid finite size effects. The relaxation time was $10^4$ full Monte Carlo steps, whereby the final fraction of cooperators $\rho_C$ was then determined in the stationary state by averaging over time for another $2 \cdot 10^4$ full Monte Carlo steps. To obtain the requested accuracy for invasion rates we averaged our results by using 1000 independent runs for each parameter values.

We have also verified that the presented results are robust to variations of the interaction lattice, for example by using random graphs and small-world networks. Regardless of the properties of the underlying interaction structure among players, we have always observed qualitatively the same results.

\clearpage

\bibliographystyle{nature}

\clearpage

\noindent \textbf{Acknowledgments} \\
This research was supported by the Hungarian National Research Fund (Grant K-120785) and the Slovenian Research Agency (Grants J1-7009, J1-9112 and P5-0027). We gratefully acknowledge computational resources provided by NIIF Hungary.

\noindent \\ \textbf{Author contributions} \\
Zsuzsa Danku, Matja{\v z} Perc and Attila Szolnoki designed and performed the research as well as wrote the paper.

\noindent \\ \textbf{Competing financial interests} \\
The authors declare no competing financial interests.

\clearpage

\clearpage

\clearpage

\clearpage

\clearpage

\clearpage

\end{document}